\documentclass[twocolumn,trackchanges]{aastex701}
\usepackage{amsmath,mathtools}

\usepackage{xcolor}%

\begin{document}

\title{Suppression of Radiative Cooling in Galaxy Cluster Cores \\ 
by the Combination of AGN Heating and Sloshing}

\author[orcid=0000-0003-0058-9719,sname='Fujita']{Yutaka Fujita}
\affiliation{Department of Physics, Graduate School of Science,
  Tokyo Metropolitan University,\\
  1-1 Minami-Osawa, Hachioji-shi, Tokyo 192-0397, Japan}
\email[show]{y-fujita@tmu.ac.jp}

\author[orcid=0000-0002-8125-4509,sname='Matsumoto']{Tomoaki Matsumoto}
\affiliation{Faculty of Humanity and Environment, Hosei University, Fujimi, Chiyoda-ku, Tokyo 102-8160, Japan}
\email[show]{matsu@hosei.ac.jp}

\author[orcid=/0000-0002-8779-8486,sname='Wada']{Keiichi Wada}
\affiliation{Kagoshima University, Graduate School of Science an
d Engineering, Kagoshima 890-0065, Japan}
\affiliation{Ehime University, Research Center for Space and Cosmic Evolution, Matsuyama 790-8577, Japan}
\affiliation{Hokkaido University, Faculty of Science, Sapporo 060-0810, Japan}
\email[show]{wada@astrophysics.jp}

\begin{abstract}
Recent XRISM observations suggest that gas mixing induced by sloshing contributes to core heating.
We systematically investigate the suppression of cooling flows in galaxy cluster cool cores through three-dimensional hydrodynamic simulations that incorporate both sloshing-driven turbulence and active galactic nucleus (AGN) heating. The AGN heating is modeled as thermal energy input that mimics cosmic-ray heating. Sloshing is represented by simple waves with amplitudes $\alpha = 0$, $0.15$, and $0.3$ times the sound speed and wavelengths $\lambda = 200$, $1000$, and $2000$~kpc. We evolve each model from an isothermal initial condition to $t = 8$~Gyr.
Without AGN heating, sloshing suppresses cooling, but it cannot stop it completely unless the core is fully disrupted. Longer wavelengths promote deeper mixing and greater suppression. Sloshing can cause cooler gas to move more quickly than hotter gas. This phenomenon has been observed in a few clusters by XRISM.
When AGN heating is included, the dense central gas is heated efficiently, substantially delaying or preventing the onset of a cooling flow. However, for intermediate wave lengths, sloshing can displace the densest gas away from the AGN heating zone, reducing the feedback effect and paradoxically enhancing net cooling relative to the wave-free case.
These results highlight a non-trivial coupling between sloshing and AGN
feedback, with implications for interpreting XRISM velocity and temperature maps of cool-core clusters.
\end{abstract}

\keywords{%
  \uat{Galaxy clusters}{584} ---
  \uat{Intracluster medium}{858} ---
  \uat{Active galactic nuclei}{16} ---
  \uat{Hydrodynamical simulations}{767} ---
  \uat{X-ray astronomy}{1810}
}

\section{Introduction}
\label{sec:intro}

Galaxy clusters harbor vast reservoirs of hot, X-ray-emitting intracluster
medium (ICM) at temperatures of ${\sim}2$--10~keV. In many clusters the radiative cooling time of the core gas falls well below the Hubble time, yet deep spectroscopic observations with XMM-Newton established that the actual mass deposition rate is far smaller than the classical cooling-flow prediction
\citep{2001A&A...365L.104P, 2001A&A...365L..99K, 2001A&A...365L..87T}.
Some heating mechanism must compensate the radiative losses of the core.

Among the proposed heating mechanisms, Active galactic nuclei (AGN) feedback has attracted the most attention \citep{2007ARA&A..45..117M, 2012ARA&A..50..455F}. Bubbles of relativistic plasma inflated by the central radio galaxy have been observed in many cool-core clusters
\citep{1993MNRAS.264L..25B, 2000MNRAS.318L..65F, 2000ApJ...534L.135M, 2001ApJ...558L..15B}.
However, the details of how bubbles or AGN energy couples to the ambient ICM---through weak shocks \citep{2003MNRAS.344L..43F,2007ApJ...665.1057F,2016MNRAS.457...82S}, sound waves (\citealt{2003MNRAS.344L..43F,2006MNRAS.366..417F}, but see \citealt{2005ApJ...630L...1F,2006ApJ...638..659M}), turbulent mixing \citep{2017MNRAS.466L..39H,2020ApJ...896..104H}, turbulent dissipation \citep{2005ApJ...622..205D,2014Natur.515...85Z,2020MNRAS.494.5507F}, internal waves \citep{2018MNRAS.478.4785Z}, or
cosmic-ray heating \citep{1991ApJ...382..416B,2008MNRAS.384..251G,2013MNRAS.428..599F}---remain under active investigation \citep{2019SSRv..215....5W}.

An alternative, or complementary, heating channel is turbulence driven by
large-scale gas sloshing.
Gas sloshing is induced when the ICM is perturbed by minor mergers or off-axis passages of substructures, setting the core gas into oscillatory motion and producing the characteristic spiral-shaped cold fronts observed in many cool-core clusters
\citep{2006ApJ...650..102A}.
The sloshing motions generate subsonic turbulence throughout the core. The subsequent dissipation of turbulent kinetic energy provides an additional, spatially distributed heat source that can offset the radiative cooling of the ICM, at least in part \citep{2004ApJ...612L...9F,2010ApJ...717..908Z,2020MNRAS.494.5507F}. If the turbulence-generated eddies are large and powerful enough, they will circulate gas throughout three regions: the vicinity of the AGN, the interior of the cool core, and the exterior of the cool core \citep{2020MNRAS.494.5507F}. Thus, turbulence in the core is expected to play a dual role in the thermal energy budget: it can transport heat inward from the surrounding regions into the core, while simultaneously redistributing the thermal energy injected by the central AGN outward across the core volume.
However, it is unclear whether sloshing-driven turbulent mixing can penetrate deeply enough into the innermost, high-density region to stop cooling completely.
This inherent limitation underscores the importance of a localized heating source, such as AGN feedback, that can directly target the central density peak and maintain long-term thermal balance in cool cores.

The commissioning of the XRISM/Resolve spectrometer has opened a
new window on ICM kinematics. Observations of the Centaurus cluster revealed bulk gas motions with velocities of 130--$310\:\rm km\: s^{-1}$ in the cool core. This is the first kinematic discovery of sloshing and suggests that sloshing-driven mixing contributes to core heating \citep{2025Natur.638..365X}.
In Perseus, XRISM has disentangled multiple kinematic
drivers---including AGN-driven outflows and sloshing \citep{2026Natur.650..309T,2026A&A...707A.109Z}---and has found that
cooler gas components move at systematically higher velocities than the
hotter ambient ICM
\citep{2026arXiv260422975M}.
This kinematic signature of fast-moving cool gas could be a consequence
of sloshing, in which low-entropy gas is entrained and displaced by
large-scale eddies; testing whether our simulations can reproduce this
behavior is one of the primary observational benchmarks of the present
work.
These two drivers---sloshing and AGN activity---are thus likely to coexist in real cool-core clusters.
Previous studies that incorporated both
relied on cosmological simulations in which the parameters of cluster
mergers (e.g., mass ratio) were not well controlled
\citep{2017ApJ...849...54L,2021MNRAS.504.3922C}, and the dependence of
the results on these parameters has not been thoroughly investigated.

The interplay between sloshing and AGN feedback is likely more complex than the superposition of two independent heating channels. The efficiency of AGN heating depends critically on the spatial distribution of gas in the core. Sloshing could change the structure of
the high-density gas, and therefore the onset of a cooling flow may be
different from the AGN-only case. Furthermore, whether sloshing aids or hinders AGN feedback depends on the spatial scale of the perturbations.
 A central question of this paper is therefore to identify the boundary in sloshing parameter space—wavelength $\lambda$ and amplitude $\alpha$—that separates the regime in which sloshing assists AGN feedback from the regime in which it undermines it.

In this paper, we extend the two-dimensional sloshing wave model of \citet{2004ApJ...612L...9F}  and present three-dimensional hydrodynamic simulations that combine sloshing-driven turbulence with a physically motivated, self-regulated AGN heating prescription. We systematically vary the sloshing parameters (wave amplitude and wavelength) and the AGN feedback timescale. We then track the radiative cooling of the ICM and the time at which a cooling flow begins. Because our simulations are idealized, we focus on parameter dependence rather than detailed comparisons with observations (see \citealt{2025ApJ...993L..11X,2025arXiv251212754B}). Section~\ref{sec:model} describes the numerical model. Results are presented in Section~\ref{sec:results} and discussed in Section~\ref{sec:discussion}. Conclusions are summarized in Section~\ref{sec:conclusions}.

\section{Model}
\label{sec:model}

\subsection{Hydrodynamics and Initial Conditions}
\label{subsec:hydro}

We solve the three-dimensional Euler equations with radiative cooling
using a 3-level nested-grid code based on the SFUMATO code \citep{2007PASJ...59..905M}. Within the uniform parent grids defined for $-368\leq x,y,z \leq +368$~kpc, smaller uniform sub-grids are defined for $-184\leq x,y,z \leq +184$~kpc and $-92\leq x,y,z \leq +92$~kpc. There are $128^3$ grid points at each level. The finest cell size is 1.4~kpc. Free-boundary conditions are adopted except for the surface of wave injection (see Section~\ref{subsec:waves}), and fixed-boundary conditions are adopted when waves are not injected. The shock-stable Roe scheme \citep{2003JCoPh.185..342K} is adopted for the Riemann solver. The solver attains second-order accuracy in time and space by employing the MUSCL method and the predictor-corrector method, respectively.

The fixed gravitational potential and initial gas distribution are identical to those in \citet{2004ApJ...612L...9F,2004ApJ...600..650F}, which model a typical middle-mass cluster. We adopt the NFW profile \citep{1997ApJ...490..493N} for the mass distribution of a cluster. The mass profile is written as follows:
\begin{equation}
\label{eq:NFW}
 M(r) \propto \left[ \ln\left(1 + \frac{r}{r_s}\right) - \frac{r}{r_s\left(1 + r/r_s\right)} \right]\:,
\end{equation}
where $r_s$ is the characteristic radius of the cluster. The normalization can be expressed as $M(r_{\rm vir}) = M_{\rm vir}$, where $r_{\rm vir}$ and $M_{\rm vir}$ are the virial radius and mass of the cluster, respectively. We assume $r_s=460$~kpc, $r_{\rm vir}/r_s=4.2$, and $M_{\rm vir}=1.2\times 10^{15}\: M_\odot$ and ignore the self-gravity of the ICM. 
The ICM is initially ($t=0$) in pressure equilibrium and isothermal. We chose these conditions because they are simple and clear. For the NFW profile (equation~(\ref{eq:NFW})), the initial ICM density profile can be written as
\begin{equation}
 \rho(r) = \rho_0 \exp[-B f(r/r_s)] \;,
\end{equation}
where $\rho_0$ is the central gas density, and
\begin{equation}
 B = \frac{1}{m(1)}\frac{G\mu m_p M(r_s)}{r_s k T_{\rm 0}} \;,
\end{equation}
\begin{equation}
 m(x) = \ln(1+x)-\frac{x}{1+x} \;,
\end{equation}
\begin{equation}
 f(x) = 1-\frac{1}{x}\ln(1+x) \;,
\end{equation}
where $G$ is the gravitational constant, $\mu=0.6$ is the mean molecular weight, and $m_p$ is the proton mass \citep{1998ApJ...509..544S}. We assume an initial central electron density of $n_{e0}=0.02\rm\: cm^{-3}$ and a temperature of $k T_0=7$~keV, which reproduce the density and the temperature outside of the cool core ($r\sim 150$--200~kpc) of Abell~1795 \citep{2002MNRAS.331..635E}. The central density ($n_{e0}=0.02\rm\: cm^{-3}$) is more typical of non-cool-core clusters than cool-core systems \citep{2009MNRAS.395..764S} because the isothermal condition was adopted. 

We use the cooling function from equation (6) of \citet{2002ApJ...581..223R} for a metal abundance of $0.3\,Z_\odot$.
Thermal conduction, viscosity, and magnetic fields are neglected.
We define the cooling-flow onset time, $t_{\rm cool}$, as the first epoch at which any grid cell reaches $T < 10^{-3}$~keV.
Each run evolves from the initial state to whichever is smaller: $t=t_{\rm cool}$ or 8~Gyr. The latter is adopted as a conservative upper limit for the cluster's age\footnote[1]{It is comparable to the look-back time for $z=1$ (7.7~Gyr), assuming cosmological parameters of $H_0=70\rm\: km\: s^{-1}\: Mpc^{-1}$, $\Omega_0=0.3$, and $\Lambda=0.7$.}.

\subsection{Sloshing Waves}
\label{subsec:waves}

Cosmological simulations suggest that bulk velocities of $\sim 20$--30\% of the speed of sound are typical even for relatively relaxed clusters \citep{2003ApJ...587..524N}, and large-scale bulk motions develop on scales ${\gtrsim}100$~kpc \citep{2018PASJ...70...51O}. Sloshing waves are represented by plane waves injected at the outer boundary ($x = -368$~kpc), following \citet{2004ApJ...612L...9F}. The velocity variation is given by:
\begin{equation}
  \delta v = \alpha\,c_{s0}\,\sin\!\left(\frac{2\pi c_{s0} t}{\lambda}\right),
  \label{eq:wave}
\end{equation}
where $\alpha$ is the dimensionless amplitude, $c_{s0}=1360\:\rm km\: s^{-1}$ is the initial sound speed, $\lambda$ is the wavelength. The density and pressure variations are respectively given by:
\begin{equation}
  \label{eq:drho}
 \delta \rho = \alpha\,\rho_{b0}\,\sin\!\left(\frac{2\pi c_{s0} t}{\lambda}\right),
\end{equation}
\begin{equation}
 \delta p = \alpha\gamma\,p_{b0}\,\sin\!\left(\frac{2\pi c_{s0} t}{\lambda}\right),
\end{equation}
where $\rho_{b0}$ and $p_{b0}$ are the initial density and pressure at $x = -368$~kpc, and $\gamma=5/3$.
We explore $\alpha \in \{0,\,0.15,\,0.3\}$ and $\lambda \in \{200,\,1000,\,2000\}$~kpc (Table~\ref{tab:models}). To suppress artificial symmetry structures in the three-dimensional Cartesian grid $(x ,y, z)$, the waves in the half-space $y > 0$ are phase-shifted by $\lambda/4$ relative to those in $y < 0$. The aim of the long-wavelength model calculations ($\lambda\geq 1000$~kpc) is to understand the impact of slow environmental changes on the core ($r\lesssim 100$~kpc), not to discuss the cluster-scale structure ($\sim 1$--2~Mpc). In other words, we focus on how the period of the perturbation, rather than the wavelength, affects the results. Since we applied free-boundary conditions to all boundaries except the one through which the wave enters, the waves pass through the boundaries. We confirmed that there are no significant reflections.

\subsection{AGN Heating}
\label{subsec:agn}

Due to the low velocity dispersion observed in cluster cores by Hitomi \citep{2016Natur.535..117H} and XRISM \citep{2025Natur.638..365X,2025ApJ...982L...5X,2025PASJ...77S.270F}, we model AGN feedback as thermal energy from cosmic rays rather than as kinetic energy from AGN jets.
Rather than explicitly calculating cosmic-ray streaming, we consider a volumetric thermal energy source with a radial profile proportional to $r^{-1}$, which is motivated by the expected energy deposition profile from cosmic-ray heating \citep{2008MNRAS.384..251G,2011ApJ...738..182F,2013MNRAS.428..599F}:
\begin{equation}
  \label{eq:agn_profile}
  \dot{q}_{\rm AGN}(r) \propto 
  \begin{dcases*}
    \frac{Q_{\rm AGN}(t)}{r_{\rm in}}  & for $r<r_{\rm in}$, \\
    \frac{Q_{\rm AGN}(t)}{r}  & for $r_{\rm in}<r<r_{\rm out}$, \\
    0       & for $r>r_{\rm out}$,
  \end{dcases*}
\end{equation}
where $Q_{\rm AGN}(t)$ is the total heating rate.  We introduce the first line to prevent divergence at $r=0$ and assume $r_{\rm in}=3$~kpc, considering the resolution of the simulations. The outer radius is set to $r_{\rm out}=60$~kpc for $xy>0$ and $r_{\rm out}=30$~kpc for $xy<0$. This is because the heating is likely to be asymmetric if cosmic rays are accelerated by the jets or accretion disk of the AGN. This asymmetry also prevents the formation of artificial symmetrical structures.
The heating term is normalized so that $\int \dot{q}_{\rm AGN}\,dV = Q_{\rm AGN}(t)$. 
The outer radius of $r_{\rm out}=30$--60~kpc approximately corresponds to the size of the region influenced by the AGN for the Perseus cluster \citep{2026Natur.650..309T}.
The radial profile of the heating rate affects the results. For instance, using $r^{-2}$ instead of $r^{-1}$ in equation (\ref{eq:agn_profile}) often yields an ICM temperature of $\gtrsim 10$~keV at $r\lesssim 20$~kpc, which is generally inconsistent with observations \citep{2009ApJS..182...12C}.

Cold molecular gas has been discovered around the centers of galaxy clusters \citep{2019A&A...631A..22O,2019MNRAS.490.3025R}. This suggests that central AGNs are fueled by the cold gas rather than the hot ICM \citep{2022ApJ...924...24F}. Indeed, AGN activity correlates with the mass of the cold gas around AGNs \citep{2023PASJ...75..925F,2024ApJ...964...29F,2026ApJ..1000..256C}. The cold gas likely forms through thermal instability on small scales, even without significant mass dropout associated with strong cooling flows \citep{2012ApJ...746...94G,2012MNRAS.419.3319M,2015ApJ...811...73L,2015ApJ...811..108P,2015ApJ...799L...1V}. We assume that AGN power is proportional to the mass of the ICM with $< 3$~keV within $r_{\rm in}$. This assumption is based on observations showing that cold gas is associated with low-entropy ICM, and that the mass of cold gas correlates with the mass of the ICM \citep{2019A&A...631A..22O,2019MNRAS.490.3025R}. 
Thus, the total heating rate $Q_{\rm AGN}$ is represented by
\begin{equation}
  Q_{\rm AGN}(t) = \frac{M_{\rm hot}(<r_{\rm in},\,t)c^2}{T_{\rm flow}},
  \label{eq:agn_norm}
\end{equation}
where $M_{\rm hot}(<r_{\rm in},\,t)$ is the total mass of the hot ICM with $< 3$~keV within $r_{\rm in}$ at time $t$, $c$ is the speed of light, and $T_{\rm flow}$ is a free parameter representing the gas accretion timescale. We consider $T_{\rm flow} = 10$ and $30$~Gyr, as well as the AGN-free case ($Q_{\rm AGN} = 0$).
We note that most of the cold gas is expected to be consumed in star formation, with only a fraction flowing into the black hole \citep{2022ApJ...924...24F}.

\begin{deluxetable*}{ccccl}
\tablecaption{Simulation Parameters and Results \label{tab:models}}
\tablewidth{0pt}
\tablehead{
  \colhead{$T_{\rm flow}$} &
  \colhead{$\alpha$} &
  \colhead{$\lambda$} &
  \colhead{$t_{\rm cool}$} &
  \colhead{Notes}\\
  \colhead{(Gyr)} &
  \colhead{} &
  \colhead{(kpc)} &
  \colhead{(Gyr)} &
  \colhead{}
}
\startdata
\multicolumn{5}{c}{No AGN ($Q_{\rm AGN}=0$)} \\
\hline
No AGN & 0    & ---  & 2.7 & wave-free \\
No AGN & 0.15 & 200  & 2.7 & \\
No AGN & 0.15 & 1000 & 3.9 & \\
No AGN & 0.15 & 2000 & 4.7 & \\
No AGN & 0.3  & 200  & 3.3 & \\
No AGN & 0.3  & 1000 & 4.3 & \\
No AGN & 0.3  & 2000 & $>8$ & core destroyed \\
\hline
\multicolumn{5}{c}{$T_{\rm flow} = 30$~Gyr} \\
\hline
30 & 0    & ---  & 7.2 & wave-free \\
30 & 0.15 & 200  & 7.5 & \\
30 & 0.15 & 1000 & 4.1 & enhanced cooling$^a$ \\
30 & 0.15 & 2000 & $>8$ & core destroyed \\
30 & 0.3  & 200  & 7.6 & \\
30 & 0.3  & 1000 & 5.0 & enhanced cooling$^a$ \\
30 & 0.3  & 2000 & $>8$ & core destroyed \\
\hline
\multicolumn{5}{c}{$T_{\rm flow} = 10$~Gyr} \\
\hline
10 & 0    & ---  & 7.3 & wave-free \\
10 & 0.15 & 200  & $>8$ & \\
10 & 0.15 & 1000 & 4.1 & enhanced cooling$^a$ \\
10 & 0.15 & 2000 & $>8$ & core destroyed \\
10 & 0.3  & 200  & 7.5 & \\
10 & 0.3  & 1000 & 5.3 & enhanced cooling$^a$ \\
10 & 0.3  & 2000 & $>8$ & core destroyed \\
\enddata
\tablenotetext{a}{The cooling time in these runs is shorter than in the
  wave-free run at the same $T_{\rm flow}$, indicating that sloshing
  enhances net cooling; see Section~\ref{subsec:agn_results}.}
\end{deluxetable*}

\begin{figure*}[t]
\plotone{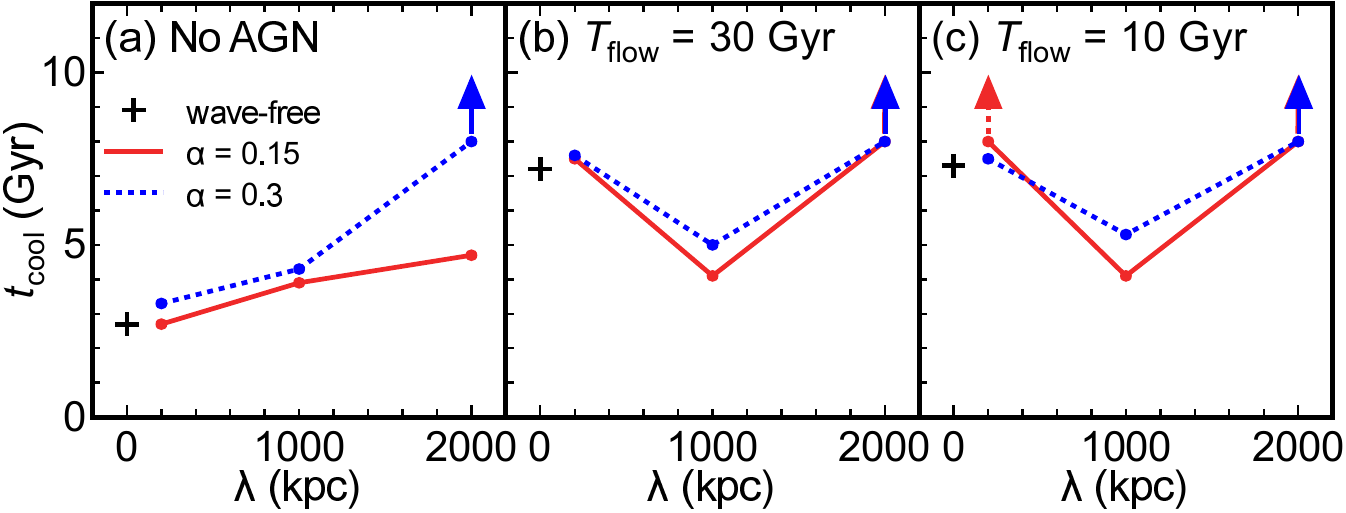}
\caption{Relation between $\lambda$ and $t_{\rm cool}$. (a) No AGN runs, (b) $T_{\rm flow} = 30$~Gyr, and (c) $T_{\rm flow} = 10$~Gyr. The crosses show wave-free runs ($\alpha=0$). The red solid lines represent $\alpha=0.15$, and the blue dotted lines represent $\alpha=0.3$. The arrows indicate $t_{\rm cool} > 8$~Gyr. The solid arrows show that the cool core is destroyed, and the dotted arrow shows that it survives.
\label{fig:tcool}}
\end{figure*}
\section{Results}
\label{sec:results}

Table~\ref{tab:models} lists all simulation runs grouped by gas accretion time scale, $T_{\rm flow}$. For each run, we record the cooling-flow onset time, $t_{\rm cool}$, as the first epoch at which any grid cell reaches $T < 10^{-3}$~keV. Figure~\ref{fig:tcool} shows the relation between $\lambda$ and $t_{\rm cool}$. An earlier $t_{\rm cool}$ indicate more efficient cooling. Wave-free runs ($\alpha = 0$) serve as baseline; varying $\lambda$ at $\alpha = 0$ is meaningless by construction.

\subsection{Sloshing Without AGN Heating}
\label{subsec:no_agn}

\begin{figure*}[t]
\plottwo{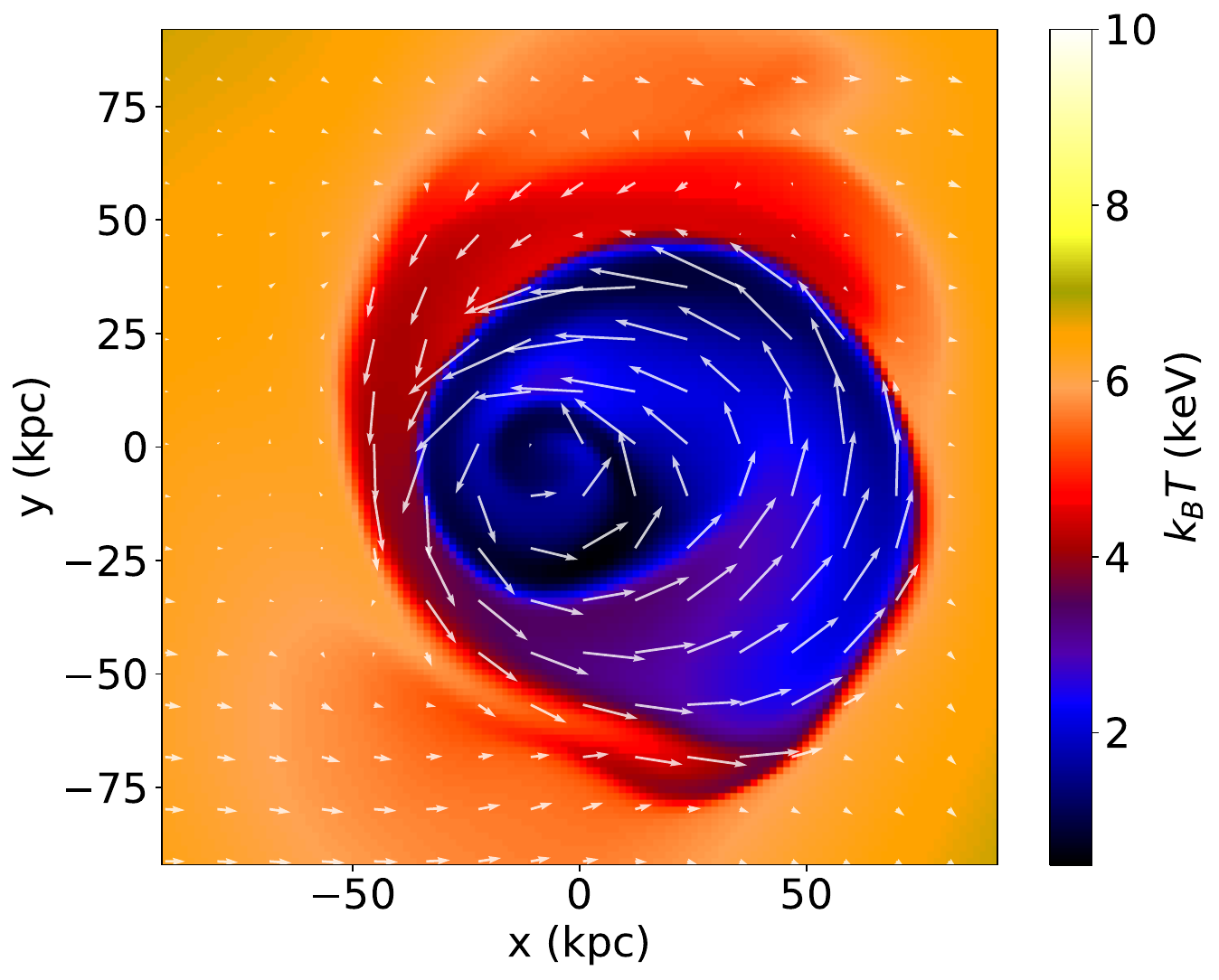}{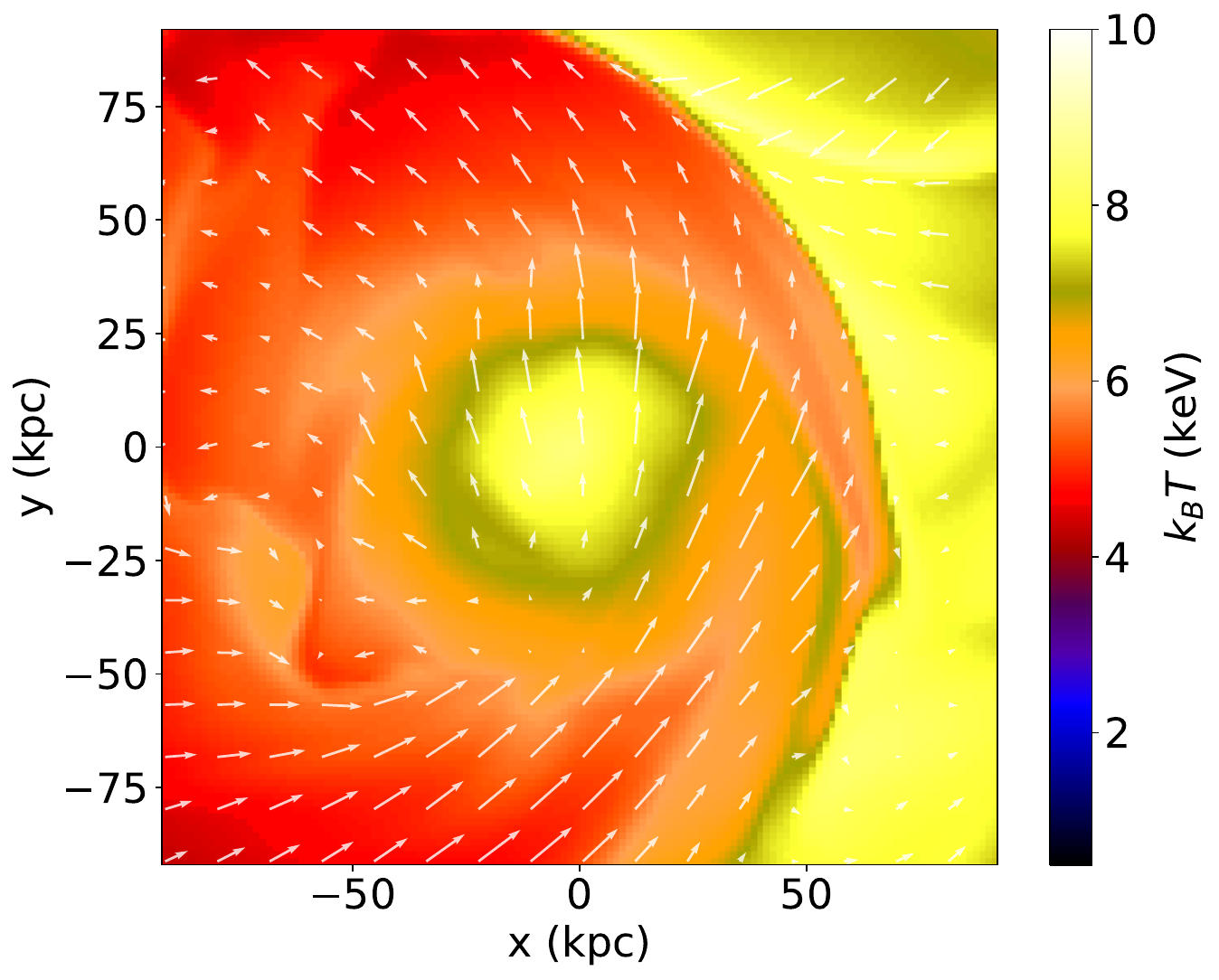}
\caption{Temperature distribution of the ICM at $z=0$ without AGN heating. The arrows show the motion of the ICM. Left: $\alpha=0.15$, $\lambda=1000$~kpc, and $t=3.9$~Gyr. Right: $\alpha=0.3$, $\lambda=2000$~kpc, and $t=6.0$~Gyr.
\label{fig:T_noAGN}}
\end{figure*}

In runs without AGN heating, waves propagate through the ICM and agitate the cool core. Because it takes $\sim 0.5$~Gyr for the wave to pass through the computational domain ($\sim 700$~kpc), the timescale for initiating the sloshing motion is shorter than the cooling time $t_{\rm cool}$ (Table~\ref{tab:models}). After a few gigayears, the sloshing motion of the waves creates turbulence within and around the core (Figure~\ref{fig:T_noAGN} left). Unlike the results of \citet{2004ApJ...612L...9F}, rotational motion and spiral structures are created due to the waves' asymmetry at $y=0$ (see section~\ref{subsec:waves}). The denser, cooler gas ($\sim 2$~keV) moves rapidly and distinctly from the surrounding, hotter gas ($\sim 5$~keV) and appears to resonate with the waves. In fact, the dynamical timescale is $t_{\rm dyn}=2\:\pi r/v_R$, where $v_R=\sqrt{G M(r)/r}$. It is $\sim 5\times 10^8$~yr at $r=75$~kpc. This is comparable to the wave period, $\lambda/c_{s0}\sim 7\times 10^8$~yr, when $\lambda=1000$~kpc. Thus, sloshing can cause the cooler gas to move quickly, as has been observed in a few clusters by XRISM \citep{2026ApJ...998..210X,2026ApJ...998..160M,2026arXiv260422975M}. This tendency was suggested by simulations in \citet{2016ApJ...821....6Z}, although those simulations did not include radiative cooling and the core temperature was higher.

Turbulent eddies generated by sloshing suppress cooling relative to the wave-free, no-AGN baseline ($t_{\rm cool} = 2.7$~Gyr), but do not halt it completely unless the core is totally disrupted (Table~\ref{tab:models} and Figure~\ref{fig:tcool}(a)).
The residual cooling originates in the region of the highest density nearest the cluster center, where the high-density gas prevents the development of large, fast-rotating eddies and efficient thermal mixing.

Both larger amplitude and longer wavelength yield more effective suppression (Table~\ref{tab:models} and Figure~\ref{fig:tcool}(a)).
For $\alpha = 0.15$, increasing $\lambda$ from 200~kpc to 2000~kpc increases the cooling time from $t_{\rm cool} = 2.7$~Gyr to $4.7$~Gyr. The extreme combination $\alpha = 0.3$, $\lambda = 2000$~kpc destroys the cool core entirely (Figure~\ref{fig:T_noAGN} right), leaving no cool gas by $t = 8$~Gyr (Table~\ref{tab:models}).
Longer-wavelength waves excite larger eddies that mix a greater volume of
gas and transport more thermal energy from the surrounding ICM into the core. This suggests that long-lasting flows from cosmological filaments \citep{2021MNRAS.502..714R} could affect core evolution more than sporadic cluster mergers.

\begin{figure*}[t]
\plottwo{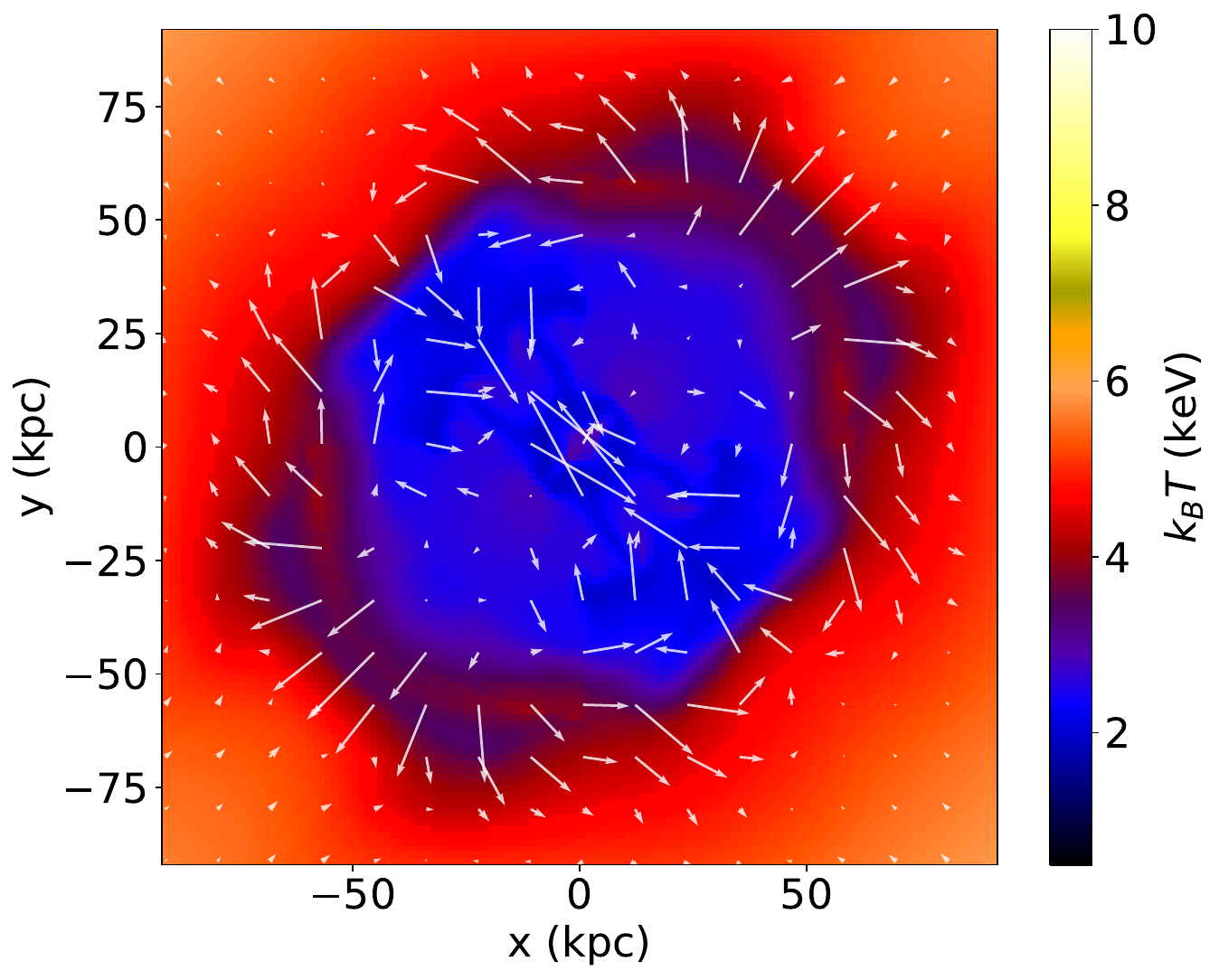}{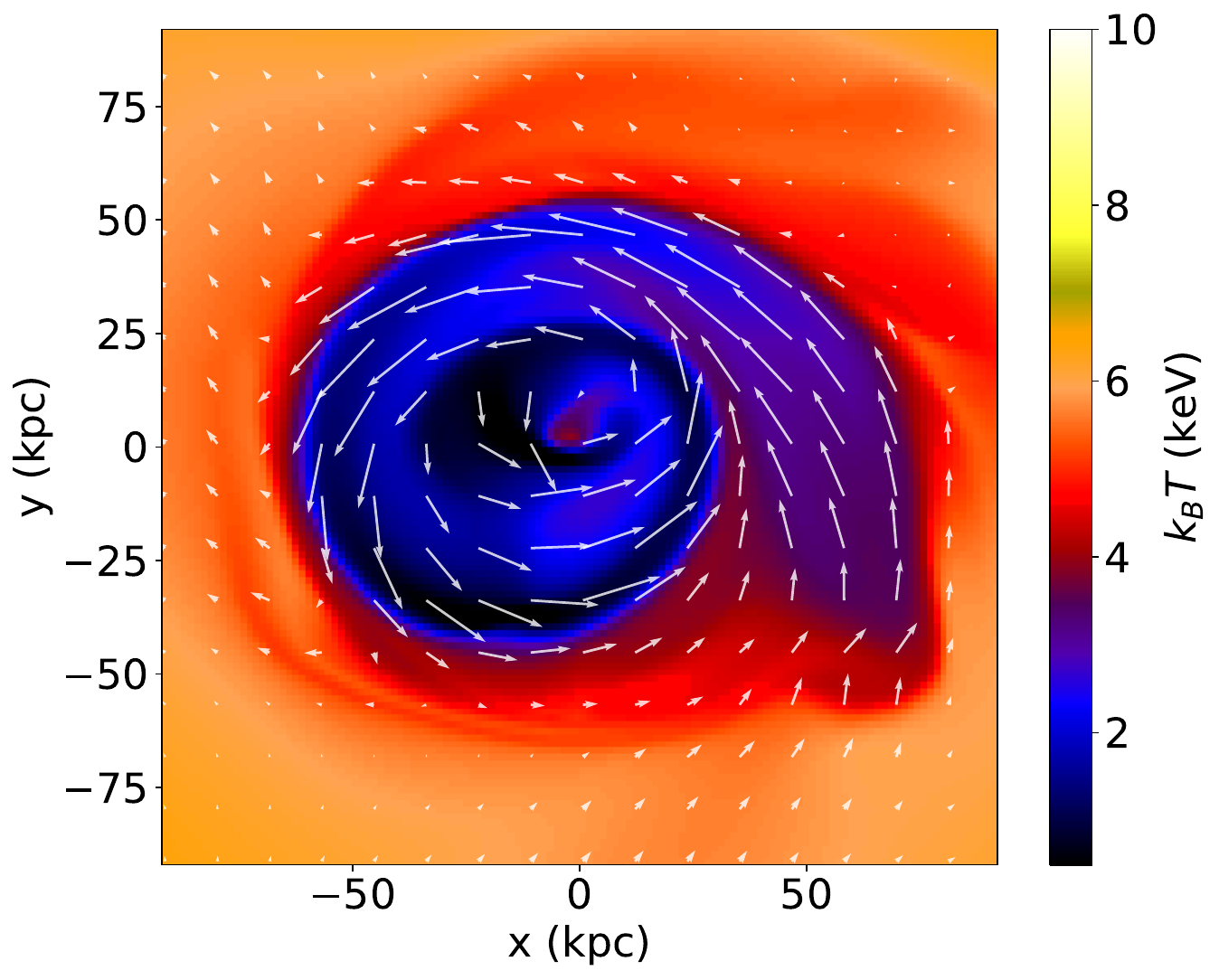}
\caption{Temperature distribution of the ICM at $z=0$ for $T_{\rm flow} = 30$~Gyr. The arrows show the motion of the ICM. Left: wave-free, and $t=7.0$~Gyr. Right: $\alpha=0.15$, $\lambda=1000$~kpc, and $t=4.1$~Gyr.
\label{fig:T_3.0}}
\end{figure*}

\begin{figure}[t]
\plotone{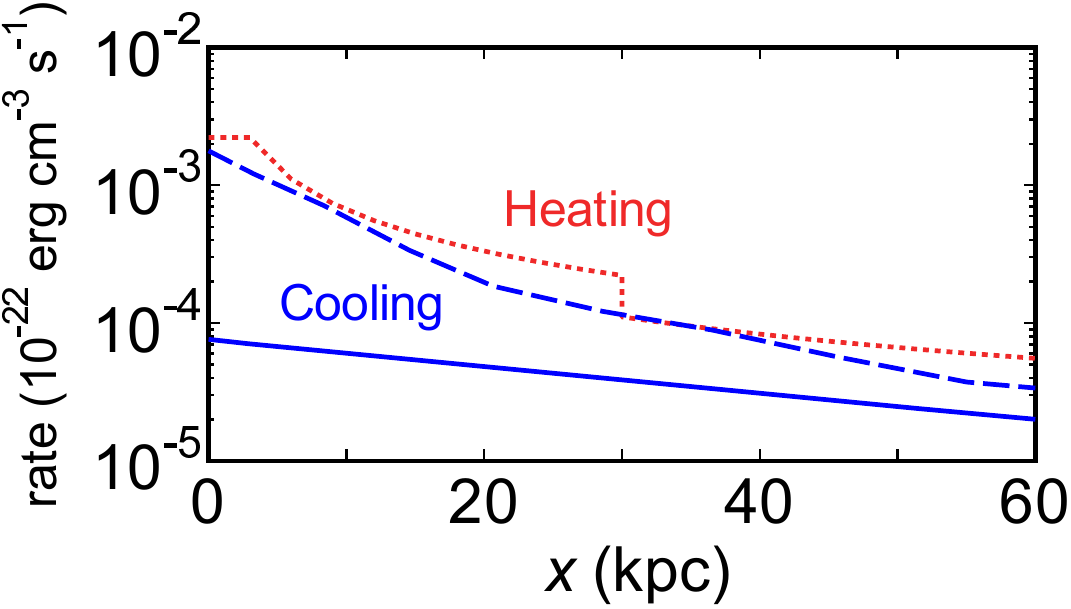}
\caption{For $T_{\rm flow} = 30$~Gyr, the wave-free run, the cooling rate ($n_e^2\Lambda(T)$) profiles are shown along the $x$-axis at $t=0$ (blue solid line) and 3~Gyr (blue dashed line). The heating rate ($\dot{q}_{\rm AGN}$) profile is also shown at $t=3$~Gyr (red dotted line).
\label{fig:h-c}}
\end{figure}

\begin{figure*}[t]
\plottwo{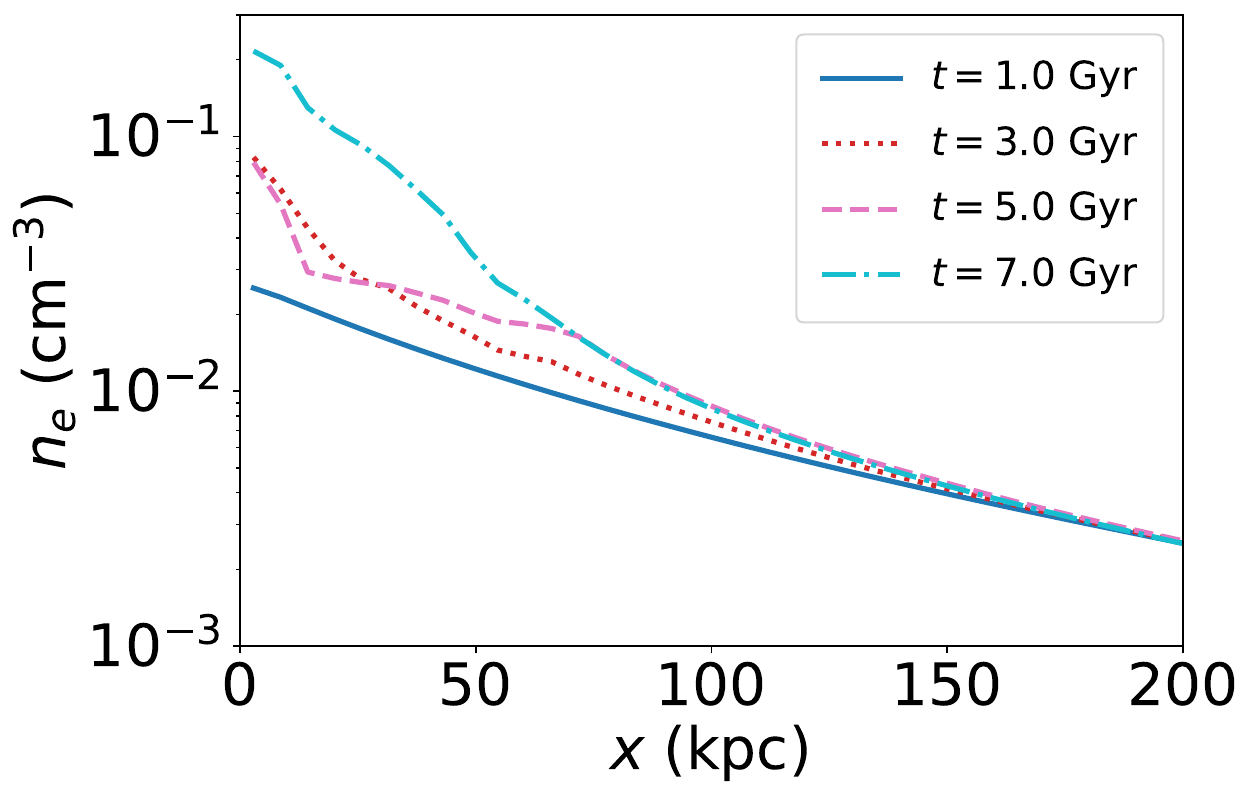}{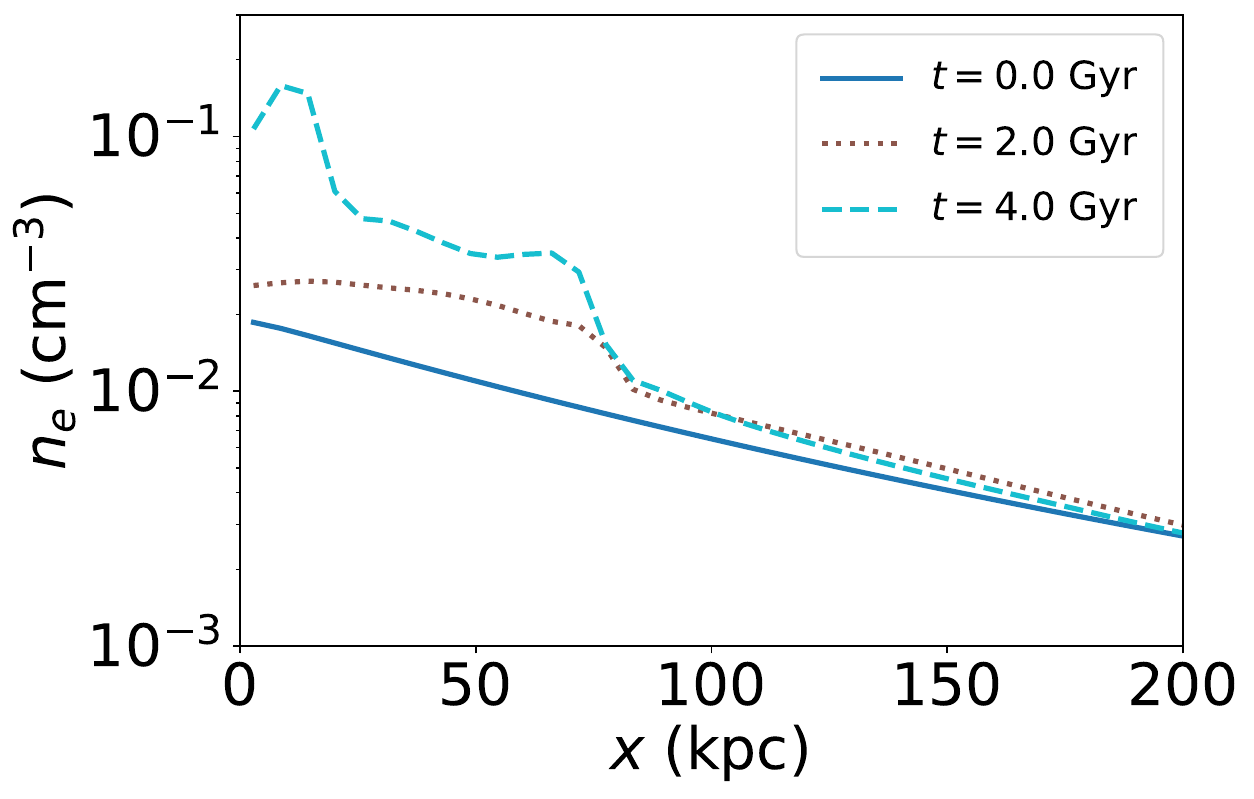}
\caption{Evolution of the density profile along the $x$-axis for $T_{\rm flow} = 30$~Gyr. Left: wave-free. Right: $\alpha=0.15$, and $\lambda=1000$~kpc.
\label{fig:T_3.0_ne}}
\end{figure*}

\begin{figure*}[t]
\plottwo{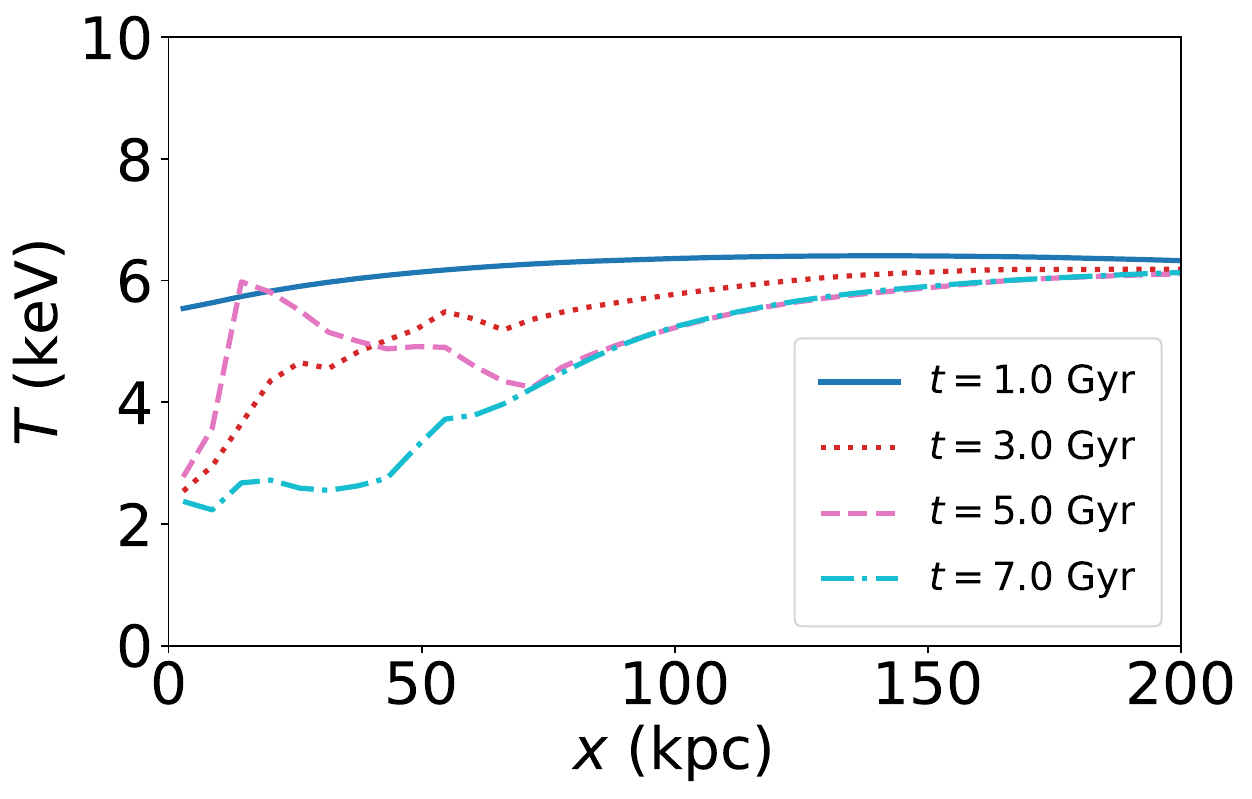}{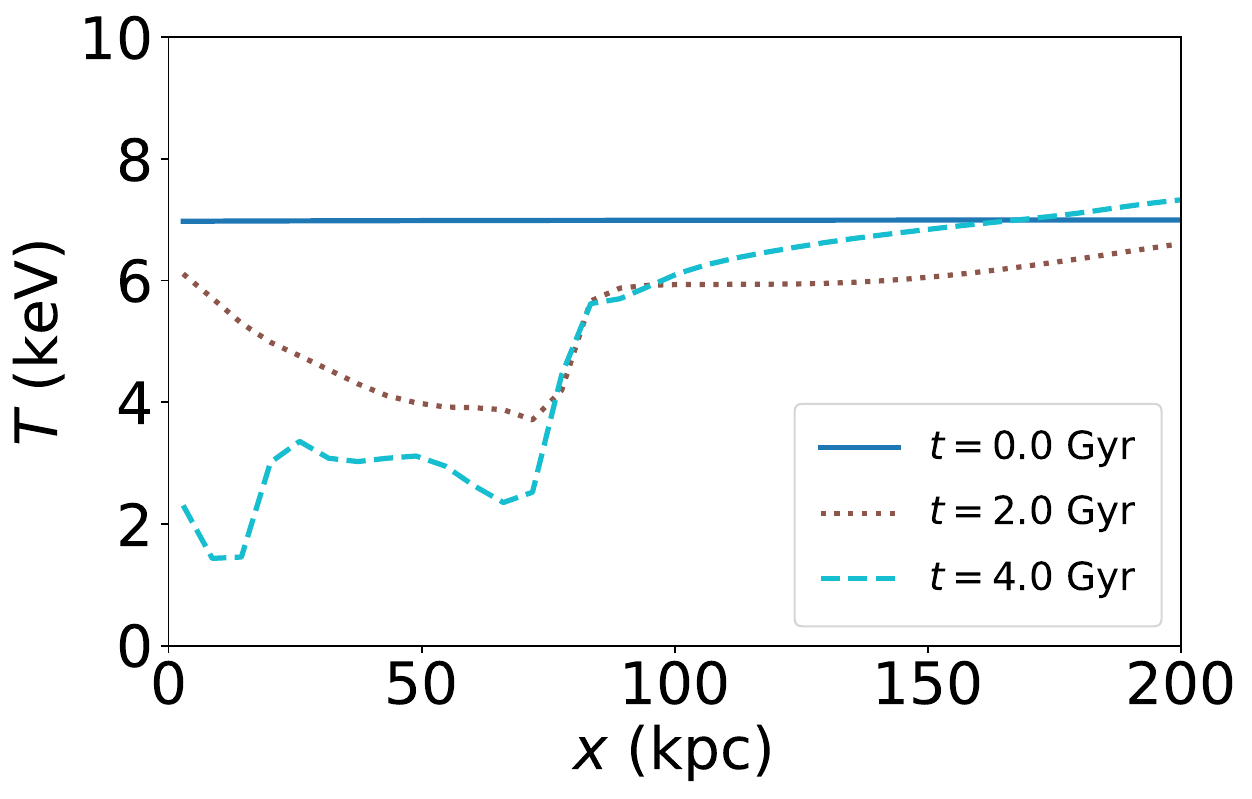}
\caption{Same as Figure~\ref{fig:T_3.0_ne} but for the temperature profile.
\label{fig:T_3.0_T}}
\end{figure*}

\subsection{Sloshing With AGN Heating}
\label{subsec:agn_results}

When AGN heating is included, the $r^{-1}$ profile (equation~(\ref{eq:agn_profile})) deposits energy preferentially in the high-density core where cooling is fastest, making feedback intrinsically efficient.
Compared to the wave-free, no-AGN baseline, the wave-free AGN runs dramatically suppresses radiative cooling (Table~\ref{tab:models}; Figures~\ref{fig:tcool} and~\ref{fig:T_3.0} left). For $T_{\rm flow} = 30$~Gyr, the wave-free run yields $t_{\rm cool} = 7.2$~Gyr. For $T_{\rm flow} = 10$~Gyr, the wave-free run gives $t_{\rm cool} = 7.3$~Gyr, compared to $t_{\rm cool} = 2.7$~Gyr for the no-AGN baseline (Table~\ref{tab:models}).
For $T_{\rm flow} = 30$~Gyr, the wave-free run, we show the profiles of the cooling rate, $n_e^2\Lambda(T)$, along the $x$-axis at $t=0$ and 3~Gyr in Figure~\ref{fig:h-c}. Here, $n_e$ is the electron density and $\Lambda(T)$ is the cooling function. The figure also shows the profile of the heating rate, $\dot{q}_{\rm AGN}$, at $t=3$~Gyr (equation~(\ref{eq:agn_profile})). Heating turns on at $t\sim 2$~Gyr when the ICM temperature around the AGN is $< 3$~keV (see Section~\ref{subsec:agn}). For $30\lesssim x \lesssim 60$~kpc, we plot half of the AGN heating rate for $xy>0$, as the AGN heating rate is zero for $xy<0$ (see Section~\ref{subsec:agn}). As shown in the figure, the cooling is almost balanced with the heating at $t=3$~Gyr.

The combination of sloshing and AGN heating is non-monotonic (Figures~\ref{fig:tcool}(b) and (c)). For $T_{\rm flow} = 30$ and $10$~Gyr and for $\alpha = 0.15$ and 0.3, the cooling time at $\lambda = 1000$~kpc ($t_{\rm cool}=4.1$--5.3~Gyr) is significantly reduced compared to the wave-free AGN run at the same $T_{\rm flow}$ ($t_{\rm cool}=7.2$--7.3~Gyr).
We attribute this counterintuitive result to gas displacement.
The innermost gas has the highest radiative cooling rate because it is the densest. Sloshing displaces the densest gas away from the center.
Due to the $r^{-1}$ heating profile, the displaced gas is heated less. Furthermore, the decreased gas density around the AGN lowers the gas accretion rate and power output of the AGN. Consequently, increased cooling offsets the additional thermal mixing provided by the waves, allowing cooling to occur faster than in a wave-free environment.

This gas displacement effect is most pronounced at intermediate wavelengths ($\lambda = 1000$~kpc; Figures~\ref{fig:tcool}(b) and (c)), where the eddies are large enough to transport the density peak from the center outward, while they are not large enough to completely destroy the core.
Figure~\ref{fig:T_3.0} (right) shows that low-temperature gas ($\lesssim 2$~keV) is distributed up to $r\sim 50$--80~kpc.  Figures~\ref{fig:T_3.0_ne} and~\ref{fig:T_3.0_T} illustrate the evolution of the density and temperature profiles along the $x$-axis for $T_{\rm flow} = 30$~Gyr. At $2\lesssim t \lesssim 4$~Gyr, the sloshing wave model exhibits jumps (cold fronts) at $x\sim 80$~kpc (Figures~\ref{fig:T_3.0_ne} and~\ref{fig:T_3.0_T} right). The profiles are flatter at $x\lesssim 80$~kpc compared to the wave-free model (Figures~\ref{fig:T_3.0_ne} and~\ref{fig:T_3.0_T} left). This indicates that the dense, cooler gas is displaced from the innermost region by the eddies created by the waves.

At short wavelengths ($\lambda = 200$~kpc) the eddies are too small
to displace gas significantly, so AGN
heating is only modestly perturbed.
At long wavelengths ($\lambda = 2000$~kpc) the turbulence is strong
enough to destroy the core entirely.

\section{Discussion}
\label{sec:discussion}

\subsection{Physical Interpretation}
\label{subsec:physics}

Our results show that three-dimensional sloshing turbulence delays, but does not permanently halt, cooling in the absence of supplementary heating unless the cool core is destroyed. The key limitation is that turbulent eddies cannot penetrate deeply enough to mix the innermost density peak. If they do, the entire core is destroyed ($\lambda =2000$~kpc in Table~\ref{tab:models} and Figure~\ref{fig:tcool}(a)). This finding is consistent with the two-dimensional results reported in \citet{2004ApJ...612L...9F} and underscores the importance of an additional heating source, such as AGNs. 

The waves could accelerate radiative cooling due to an increase in density $\delta\rho$ (equation~(\ref{eq:drho})). However, our no AGN calculations deny this (Figure~\ref{fig:tcool}(a)). In fact, since the enhancement of cooling occurs only in the presence of AGN heating and at intermediate wavelengths ($\lambda =1000$~kpc in Table~\ref{tab:models} and Figure~\ref{fig:tcool}(b) and (c)), the primary cause is the displacement of high-density gas from the centrally concentrated AGN heating region, rather than thermal instability driven by wave injection.

The gas displacement effect could impact AGN feedback. Chaotic cold accretion, for example, depends on the gas supply near the black hole \citep{2017MNRAS.466..677G}, so sloshing-induced displacements of the central gas reservoir modulate feedback independently of the accretion prescription. Importantly, the thermal balance of a cool core cannot be assessed from AGN power or sloshing amplitude alone. Their non-trivial, coupled effect on the central density distribution must be considered. 

\begin{figure*}[t]
\plottwo{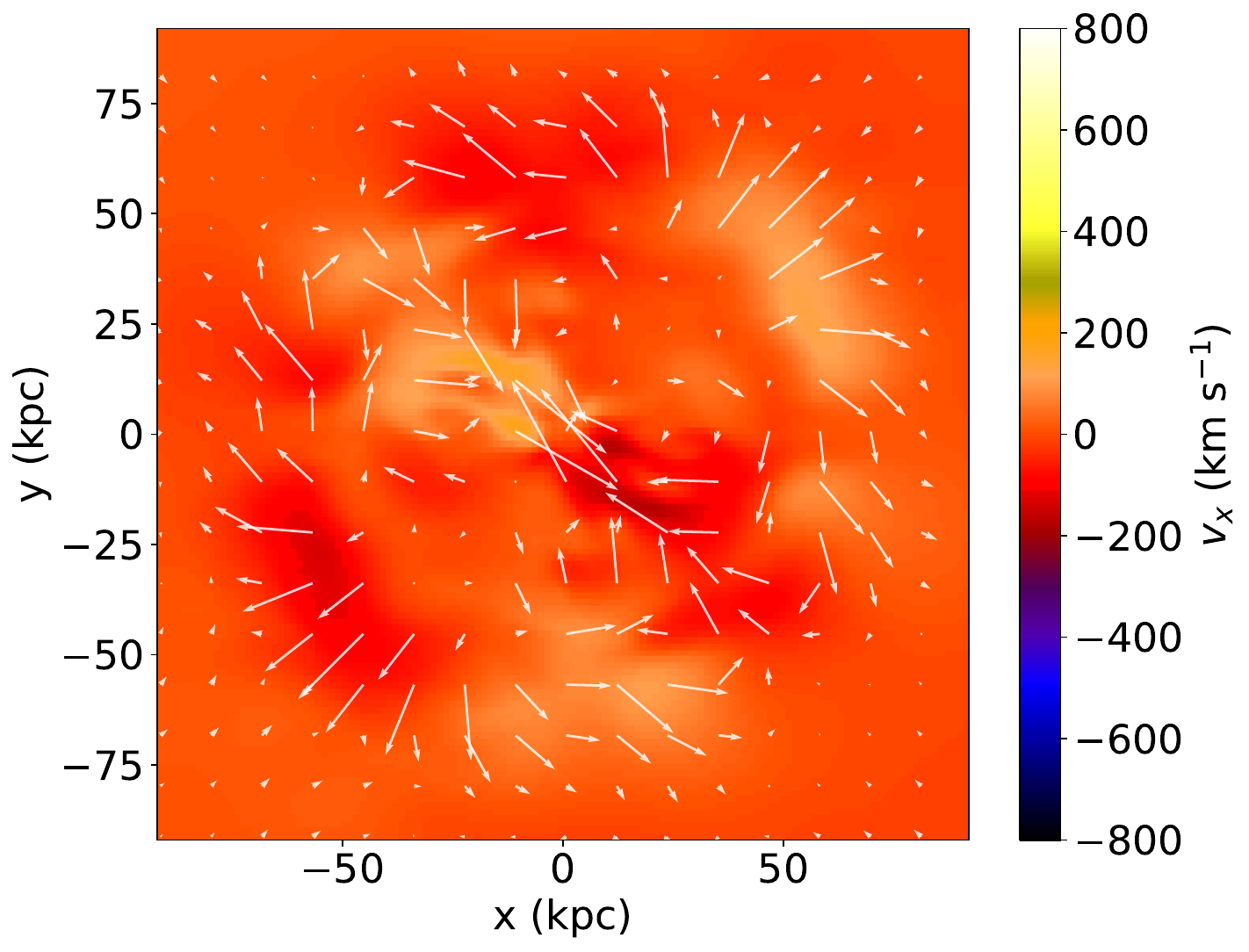}{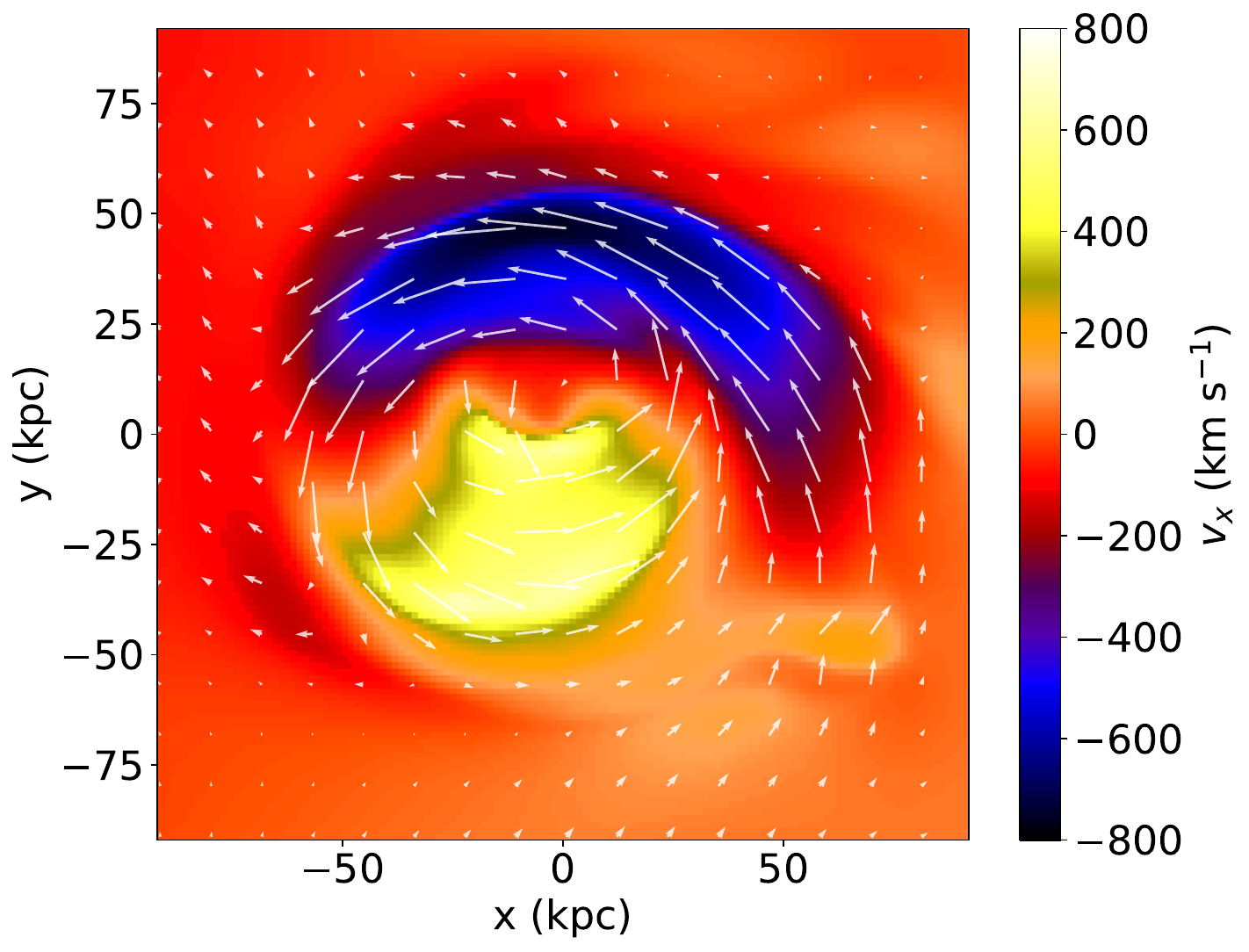}
\caption{Same as Figure~\ref{fig:T_3.0}, but for the $x$-component of velocity ($v_x$). 
\label{fig:v}}
\end{figure*}

\subsection{Spatial Resolution of the Simulations}
\label{subsec:resolution}

The spatial resolution of the simulations may affect the results. We ran a simulation with the $T_{\rm flow}=30$~Gyr, $\alpha=0.15$, and $\lambda=1000$~kpc model, using twice the resolution ($256^3$~grid points) of the original ($128^3$~grid points). We confirmed that the evolution of the former is nearly identical to that of the latter. The cooling time of the former, $t_{\rm cool} = 3.9$~Gyr, is not much different from that of the latter, $t_{\rm cool} = 4.1$~Gyr (Table~\ref{tab:models}).

\subsection{Implications for XRISM Observations}
\label{subsec:xrism}

Figure~\ref{fig:v} shows the distribution of the $x$-component of velocity ($v_x$). The figure on the left (wave-free case) shows that the velocity is fairly small ($\lesssim 150\rm\: km\: s^{-1}$) because the AGN's energy injection is in the form of thermal energy rather than kinetic energy. On the other hand, the figure on the right shows gas rotation displacing cooler gas. In fact, the high-velocity region coincides with the low-temperature region in Figure~\ref{fig:T_3.0} (right), as has been observed in a few clusters by XRISM \citep{2026ApJ...998..210X,2026ApJ...998..160M,2026arXiv260422975M}. XRISM velocity measurements will provide direct observational evidence for the physics explored here in many clusters.
Our simulations reveal that the coupling between sloshing and AGN feedback exhibits non-monotonic and parameter-dependent behavior. Future XRISM observations that jointly constrain the velocity amplitude, turbulent eddy scale, and central gas density are essential for distinguishing between sloshing- and AGN-dominated regimes.

\subsection{Caveats and Future Work}
\label{subsec:caveats}

Several simplifications limit the present models.
For example, we represent sloshing with idealized plane waves that have a fixed amplitude, wavelength, and direction. However, realistic sloshing involves a broad spectrum of perturbations from multiple directions \citep{1998Sci...280..400B,2016MNRAS.461..412Z,2025ApJ...985L..20X}.
This may lead to more complex motion of the ICM in the core.
Fixing the cluster gravitational potential also precludes the
off-center subcluster passages that drive real sloshing events.

Our AGN model assumes a smooth, time-averaged energy input, but real AGN activities are episodic. Furthermore, we did not track the evolution for $t > t_{\rm cool}$. At this stage, a large amount of cold gas flows into the AGN. This could result in powerful bursts \citep{2020MNRAS.494.5507F}, as observed in a few clusters \citep{1996A&ARv...7....1C,2005Natur.433...45M,2020ApJ...891....1G}.
Incorporating self-consistent cosmic-ray acceleration and propagation in future three-dimensional simulations will also be important.
Our model implicitly assumes that AGN feedback is always centrally peaked. This assumption is appropriate if the cosmic rays originate from the AGN's accretion disk \citep{2015ApJ...806..159K,2026PASJ...78..166S}. However, if AGN-jet-driven bubbles are the source, this assumption may be incorrect because these bubbles may be advected with the surrounding sloshing atmosphere.

Finally, although we did not consider magnetic fields, they could suppress the mixing of high- and low-entropy gases in the core \citep{2011ApJ...743...16Z}. The next step is to perform magnetohydrodynamic simulations coupled with the self-regulated AGN prescription.

\section{Conclusions}
\label{sec:conclusions}

We have presented three-dimensional hydrodynamic simulations of galaxy
cluster cool-core evolution with combined sloshing-driven turbulence
and self-regulated AGN heating.
Our main conclusions are:

\begin{enumerate}

\item \textbf{Sloshing alone cannot halt cooling.}
Sloshing delays the onset of a cooling flow, allowing cooler ICM to move faster. Larger amplitudes and longer wavelengths provide greater suppression.
However, cooling is never permanently arrested without complete core disruption because turbulent mixing cannot penetrate the innermost, high-density region.

\item \textbf{AGN feedback is highly efficient.}
The $r^{-1}$ heating profile preferentially heats the densest gas near the cluster center. This delays the onset of cooling flow by ${\gtrsim}\: 5$~Gyr compared with the unheated baseline.

\item \textbf{Sloshing can strengthen radiative cooling by displacing gas.}
For $\alpha = 0.15$--$0.3$ at $\lambda = 1000$~kpc, sloshing displaces the densest gas from the AGN vicinity. This reduces the gas accretion rate onto the AGN while enhancing the net cooling relative to the wave-free AGN case.

\item \textbf{The sloshing--AGN coupling is non-monotonic.}
The gas density field in the core must be tracked self-consistently
to determine whether sloshing aids or hinders AGN feedback.
Neither mechanism alone nor their naive superposition captures the
full thermal evolution of cool cores.

\end{enumerate}


\begin{acknowledgments}
We would like to thank the anonymous referee for their useful comments. We acknowledges support from JSPS KAKENHI Grant No.\ 23H04899, 25H00672 (Y.F.), 23K03464, 26K07154 (T.M.), 25H00671 (K,W.).
\end{acknowledgments}

\bibliography{slosh}{}
\bibliographystyle{aasjournalv7}

\end{document}